\documentclass[12pt]{article}
\usepackage{graphics}
\input epsf

\newcommand{\be}{\begin{eqnarray}}
\newcommand{\ee}{\end{eqnarray}}

\begin{document}
\title{
\begin{flushright}
{\large UAHEP043}
\end{flushright}
\vskip 1cm
A SUSY origin of gamma ray bursts}
\author{L. Clavelli\footnote{lclavell@bama.ua.edu}\\
Department of Physics and Astronomy\\
University of Alabama\\
Tuscaloosa AL 35487\\ }
\maketitle
\begin{abstract}
Bright bursts of gamma rays from outer space have been puzzling 
Astronomers for more than thirty years and there is still no
conceptually complete model for the phenomenon within the standard
model of particle physics.  Is it time to consider a supersymmetric
(SUSY) origin for these bursts to add to the astronomical indications
of supersymmetry from dark matter?
\end{abstract}
\renewcommand{\theequation}{\thesection.\arabic{equation}}
\renewcommand{\thesection}{\arabic{section}}
\section{\bf Introduction}
\setcounter{equation}{0}
   Several times each day, satellites in near earth orbit observe a
prodigious burst of gamma rays and the accumulating data has resisted
a complete physical explanation for decades.  
How long should one wait until 
deciding that it is time to consider explanations beyond the standard
model?  Astronomers, while recognizing a persistent mystery behind the
``central engine'' of violent astrophysical events, have been notoriously 
slow to encourage discussion of new physics solutions.

Ideally, in a situation like that of the gamma ray bursts (grb's), 
there should be a broad search for conceptually complete
explanations that predict the primary characteristics
of the phenomenon in zeroth order.
Violent astrophysical events are complicated processes and one must
expect difficult numerical computations to be necessary even if the underlying 
physical model is conceptually complete.  By conceptually complete 
I mean that each step of the process is completely based on known or 
plausibly proposed physical principles.  We suggest that supersymmetry 
does provide a conceptually complete picture of the burst phenomenon 
while the standard model of particle and astrophysics does not.

    Within the standard approaches to gamma ray bursts there are significant conceptual gaps.  Among these are the mechanisms for the enormous energy release required, the mechanism for the transfer of this enormous energy into a narrow range of the gamma ray spectrum, and the mechanism for the strong angular collimation of the bursts if such 
exists or the extra energy release otherwise required.  
The places where more physics input is needed are currently marked by 
the signs  ``central engine'', ``firecones'', ``sub-jets'' etc.

     In addition, the standard approaches do not predict the primary
quantitative characteristics of the gamma ray bursts except as related to
free parameters in the theory.  Most of the studies of these bursts take
as a starting point the unexplained production of relativistic outgoing jets by a dying star and proceed to model the afterglow left in their wake. Several good reviews \cite{Zhang}\cite{Rosswog} of the standard approaches are available.

\vspace {0.25in} 

     The primary characteristics of the phenomenon are:

\begin{enumerate}
\item Dominant photon energies in the $E_\gamma = 0.1$ to $1$ MeV range.
\item The distribution of burst durations ranges from approximately
$0.02$ s to $300$ s with a pronounced dip \cite{Kouveliotou} in the distribution at $2$ s.  Bursts of less than $2$ s duration are 
referred to as ``short''
bursts and those above $2$ s are defined to be ``long'' bursts.
\item Total burst energy of the order
\be
            E_T \approx 3 \cdot 10^{53} \frac{\Delta \Omega}{4 \pi} 
            {\displaystyle ergs}
\ee
where $\Delta \Omega$ is the opening solid angle of the burst.  
If $\Delta \Omega$ is $4 \pi$ (isotropic burst), the energy output is
equivalent to or greater than that observed in supernovae.  The difference, of course, is that supernovae put most of their energy release into neutrinos and into kinetic energy of heavy particles
and nuclei, eventually appearing in a broad range of the electromagnetic
spectrum.  Subject to some possibly strong assumptions, 
arguments have been given \cite{Frail} that the jet opening angles are
$\approx 5 \deg$.
This would correspond to $\Delta \Omega \approx 7.6 \cdot 10^{-3}$.
and $E_T \approx 1.8 \cdot 10^{50}$ ergs.
For comparison with the implied burst energies, the rest mass of the
sun is $1.9 \cdot 10^{54}$ ergs and that of the earth is 
$6 \cdot 10^{48}$ ergs.

\end{enumerate}

Thus, the shortest bursts can be compared to a hypothetical near solar mass of matter
disappearing into $MeV$ scale photons in $20$ ms.  Relativity would then require
 that this solar mass should be initially compressed into the radius of the
 earth ($R_E/c=.02$s) or less.  Such is the typical density of white dwarf stars.
 However, the sun is prevented from converting its full rest energy to photons
 by baryon number conservation which is well tested on time scales up to
 $10^{32}$ yr.  In addition, compact objects such as white dwarfs are
 Fermi degenerate systems
 restricted in the energy they can emit by the Pauli principle.  The electrons
 occupy all energy levels up to the Fermi energy.  The average and total 
 kinetic energy of a 
 degenerate electron gas in a white dwarf of solar mass and earth radius are, respectively $0.1$ MeV and 
 $1.2 \cdot 10^{50}$ ergs, not far from the observations quoted above assuming a $5^\circ$ opening angle.  

    This suggests that one consider the elimination of the Pauli blocking by a transition to a system of Bosons as is possible 
(perhaps even necessary) given supersymmetry and other physical considerations\cite{CK}.  
 A neutral system of density $\rho$, fully relieved of Pauli blocking, will
 collapse to a black hole in a time
\be
    t = \frac{\pi}{2}\left( \frac{8 \pi G_N \rho}{3}\right)^{-1/2} 
      \approx 1.5 s \left( \frac{\rho}{\rho_WD} \right)^{-1/2} .
\ee 
  Thus one might expect that the bursts of duration somewhat above $1.5$ s
will be depleted and might cause a dip in the duration distribution as is
observed at $2$ s.  Bursts of duration much less or much longer than this
time might be relatively unaffected by the cutoff due to gravitational 
collapse.  A full treatment, however, requires studying the collapse 
during a gradual lifting of the degeneracy pressure.

    Although a SUSY phase transition to a system of Bosons can easily explain
the primary characteristics of the gamma ray bursts as outlined below, 
my astronomer friends tell me that substantial progress is also being made
in the standard astrophysical approaches to the short duration bursts.
Among the most recent ideas that have been proposed is the following \cite{Lee}.

    Depending on various parameters, up to $10^{52}$ ergs of energy could be released into a neutrino-antineutrino cloud which could then be converted 
on a short time scale into an $e^+e^-$ cloud and made available for 
the production of a relativistic fireball.  
Assuming a one percent conversion efficiency, the fireball would have about the right rest energy, $10^{50}$ ergs, to account for the gamma ray bursts. It would seem that such standard model explanations for the bursts still do not have the predictive power or the conceptual completeness of the SUSY phase transition model.  Similarly, standard model attempts at a unified model of the bursts with a dip in the duration distribution at $2$ s often assume the dip is due to a viewing angle dependence \cite{Zhang,Yamazaki}.  
This, however, would seem to require a somewhat strange and ad hoc shape of the gamma ray wave packet.

     The SUSY phase transition model for gamma ray bursts
is based on the following scenario.

\begin{enumerate}
\item  In a region of space with a high level of fermion
degeneracy there is a phase transition to an exactly 
supersymmetric
ground state.  In the SUSY phase, electrons and their SUSY
partners (selectrons) are degenerate in mass as are the
nucleons and snucleons, photons and photinos etc.  We assume
that these masses are at most those of the standard model
particles in our normal world of broken supersymmetry.
This assumption is, perhaps, supported by the superstring
prediction of low mass (massless) ground state supermultiplets.
\item In the SUSY phase, electron pairs undergo
quasi-elastic scattering to selectron pairs 
\be
     e(p_1) + e(p_2) \rightarrow {\tilde e}(p_3) + {\tilde e}(p_4)
\label{sigma}
\ee
which, uninhibited
by the Pauli principle, can fall into the lowest energy state
via gamma emission.  These gamma rays are radiated into the
outside (non-SUSY) world.  Selectrons and other SUSY particles
are confined to the SUSY bubble being too low in energy to 
penetrate into the broken SUSY world where their mass would be 
orders of magnitude greater.  Since there are about 
\be
     N=6 \cdot 10^{56}
\ee
electrons in a white dwarf and the average squared momentum in
the degenerate sea is
\be
     <p^2> = \frac{3}{5}\left( \frac{3 N}{8 \pi V} \right)^{2/3}
        \cdot (2 \pi \hbar)^2 \approx 0.15 {\displaystyle (MeV/c)}^2,
\ee
the total energy released if all the electrons convert is
\be
    E_T = N (<E>-mc^2) = 1.2 \cdot 10^{50} {\displaystyle MeV}
\ee
as advertised above.  
\item

The highly collimated
jet structure might be produced by the stimulated emission of
sfermions and photons; in a bath of pre-existing sfermions,
new sfermion production goes preferentially into momentum
space bins with large occupation numbers.
\item
Simultaneous with electron conversion into selectrons,
nucleons within heavy nuclei convert into snucleons.
With no further support from the electron
degeneracy, the star collapses to nuclear density under
gravitational pressure.
\item  Remaining nucleon pairs then undergo the analogous
conversion to snucleon pairs with the cross section mediated
by the strong exchange of supersymmetric pions.  This process 
can be temporarily interrupted by brief periods of fusion 
energy release but then continues until all kinetic energy is
radiated away or until the star falls below 
the Schwarzschild radius and becomes a black hole.
\end{enumerate}

    In this model many of the bursts are due
to the decay of isolated white dwarfs or neutron stars which 
are absolutely table in standard astrophysics.  We therefore predict 
the existence of low mass black holes below the Chandrasekhar limit.
Other bursts might come from
the SUSY transition during a stellar contraction
when a stage of fermion degeneracy is reached.  If the duration
is linearly related to the radius of the star, the longest 
duration bursts would come from a star of radius $10^4$ times
that of earth ($10^2$ times that of the sun).  However, if the
photons undergo a random walk with a mean free path of earth
radius before emerging from the dense
star, durations would depend quadratically on radius
\be
      \tau \approx \frac{R^2}{c R_E} .
\ee
In such a case, the longest duration bursts would come from
stars of radius $100$ times that of earth.  Many other effects,
however, including fusion reignition and the free collapse
time of a star relieved of Pauli blocking, can also influence the 
duration of the burst so the situation needs more study.

\begin{figure}[ht]
\centerline{\epsfxsize= 4.1in\epsfbox{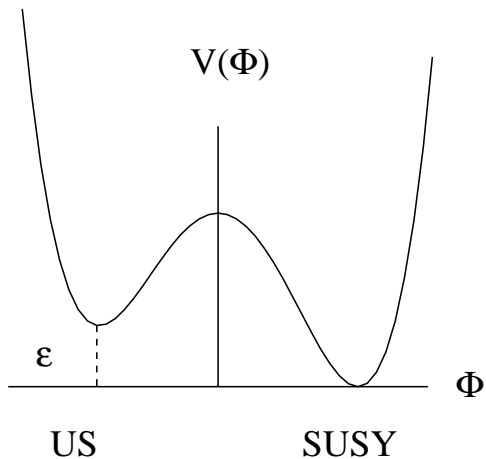}}
\caption{Effective potential in vacuum.  The broken SUSY world
is indicated with a postive vacuum energy density 
while the SUSY phase
has zero ground state energy density.
\label{fig1}}
\end{figure}

There are now accumulating indications from accelerators and 
astrophysics that we live in a broken SUSY world with a 
positive vacuum energy $\epsilon \approx 3560$ MeV/m$^3$.
On the other hand string theory seems to persistently 
predict that the true vacuum is an exact SUSY world with
zero vacuum energy.  In this situation it is inevitable
that bubbles of true vacuum will spontaneously form 
in our world and, if of greater than some critical
size, $R_c$, will expand to engulf the universe.  The potential
of some effective scalar field may be as indicated in
figure \ref{fig1} where our local minimum of broken SUSY
is labelled ``US'' and the true vacuum is exactly supersymmetric.  
In reality, $\Phi$ may stand for a complicated
linear combination of string moduli and there may be a
large number of local minima most of which are of a height
${\mathcal O}(10^{100})$ times greater than $\epsilon$.

\begin{figure}[ht]
\centerline{\epsfxsize= 4.1in\epsfbox{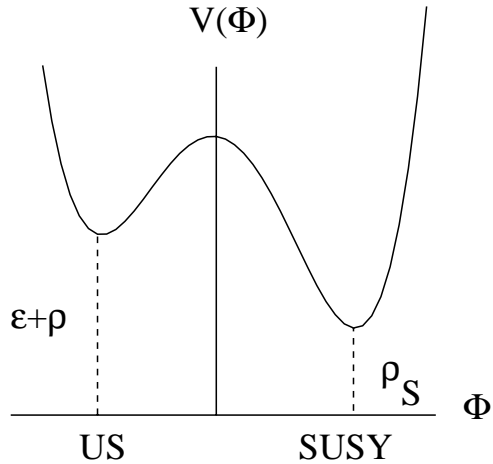}}
\caption{Effective potential in the presence of
matter.  
\label{fig2}}
\end{figure}

In the simplified single field situation the critical radius 
is obtained by an instanton calculation \cite{Coleman}.
\be
    R_c = \frac{3 S}{\epsilon} .
\label{Rc}
\ee
Here $S$ is the surface tension of the bubble of true
vacuum. Since it is to be expected that the probability to 
nucleate a bubble of radius r is a steeply falling
function of r, for sufficiently large $\frac{S}{\epsilon}$
the broken SUSY phase is metastable \cite{Frampton}
in dilute matter.
Many authors have discussed a catalysis of the phase
transition in the presence of dense matter \cite{Gorsky},
\cite{Voloshin}.  If the effective field, $\Phi$, is a
gravitational modulus, it is reasonable to assume that
the critical radius is related to the difference in
ground state energy densities 
as suggested by figure \ref{fig2}; that is
\be
     R_c = \frac{3 S}{\epsilon + \Delta \rho}
\ee

where $\Delta \rho$ is the ground state matter
density, $\rho$ in the broken SUSY phase minus the
corresponding ground state matter density, $\rho_S$
in the SUSY phase.  In a degenerate star $\Delta
\rho$ is the kinetic energy density of the 
Fermi gas since, in the SUSY phase, the ground
state will consist mainly of Bosons in the
lowest energy level.  The phase transition
to the exact SUSY vacuum is then enormously
exothermic.  For the typical white dwarf
\be
    \Delta \rho = 6 \cdot 10^{34} \displaystyle{MeV/m}^3
\ee
Thus we might reasonably expect that the ratio of the
critical radius in a white dwarf star to that in 
vacuum is
\be
    \frac{R_c(WD)}{R_c(vac)} = \frac{\epsilon}{\Delta \rho}
                     = 5 \cdot 10^{-32} .
\ee  
Here we have made the simplest assumption, namely that the
surface tension is independent of the density.  Other 
possibilities might be explored.  If the critical radius
\cite{CK}
in a white dwarf is of order $10^{-4}$ m, then the critical 
radius in vacuum is many times larger than the galactic radius.  
Thus a supercritical SUSY bubble in a
white dwarf would grow only to the edge of the star where
it would then be stabilized from further growth.  

   A transition to a Bosonic final state as proposed in this
model not only allows the release of the fermion kinetic energy
but also suggests significant jet structure due to the Bose
enhancement effect familiar from laser physics.

   In addition to the momentum dependence of the elementary 
process, the matrix element for the emission of a selectron pair with momenta $\vec{p}_3$ and $\vec{p}_4$ in process
\ref{sigma} in the presence of a bath of previously emitted pairs is
proportional to
\be
\nonumber
     {\mathcal M} \sim <n(\vec{p}_3)+1, n(\vec{p}_4)+1\mid a^\dagger(\vec{p}_3)
     a^\dagger(\vec{p}_4)\mid n(\vec{p}_3), n(\vec{p}_4)> \\
 \sim \sqrt{(n(\vec{p}_3)+1)}
     \sqrt{(n(\vec{p}_4)+1)}  .
\ee
     The cross section is, therefore, proportional to
$(n(\vec{p}_3)+1)(n(\vec{p}_4)+1)$ .

    The calculation \cite{Keung} of the process \ref{sigma} 
has been recently extended \cite{CP} to include the electron mass
effects which are important in the exact SUSY phase. 

\begin{figure}[ht]
\centerline{\epsfxsize= 4.1in\epsfbox{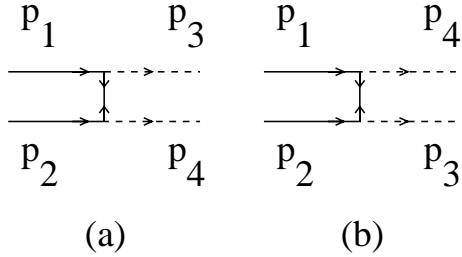}}
\caption{Feynman graphs for the conversion of an electron pair
to selectrons via (Majorana) photino exchange.
\label{fig3}}
\end{figure}

There are two
types of scalar electron (partners of the left and right handed electron)
which must be treated separately.  The three different final states are
found to have the elementary squared matrix elements
\be
\label{MELR}
   \nonumber \left|\mathcal{M}_{LR}\right|^2 =\Sigma_{ss'} \left( M_aM_a^{\dagger}+M_bM_b^{\dagger}+M_aM_b^{\dagger}+M_bM_a^{\dagger}\right)\\
   \nonumber  = e^4 \left[ \frac{1}{(t-M_{\tilde{\gamma}}^2)^2}\left(ut-2tm_e^2-\left(m_{\tilde{e}}^2-m_e^2\right)^2  \right) \right.\\ +
               \left.     t\leftrightarrow u 
       +\frac{4m_e^2\left(m_{\tilde{e}}^2-m_e^2\right)}{(t-M_{\tilde{\gamma}}^2)(u-M_{\tilde{\gamma}}^2)}\right]
\ee
where $\Sigma_{ss'}$ denotes averaging over the spins of incoming
electrons. 
\be\label{MELLRR}
  |\mathcal{M}_{RR}|^2=|\mathcal{M}_{LL}|^2=\frac{e^4M_{\tilde{\gamma}}^2}{2!}\left(s-2m_e^2\right)\left( \frac{1}{t-M_{\tilde{\gamma}}^2} + \frac{1}{u-M_{\tilde{\gamma}}^2} \right)^2.
\ee
For example the conversion rate per unit volume in the degenerate electron gas into left selectrons is given by
\be
\nonumber
      \frac{d\Gamma}{V} = \frac{{\mid \mathcal{M}_{LL} \mid}^2}{(2 \pi)^7 \sqrt{s(s-4m^2)}}
        dt \frac{{p_3}^2}{2E_3}\frac{{p_4}^2}{2E_4}dp_3 dcos(\theta_3)d\phi_3
        dp_4 dcos(\theta_4)d\phi_4 \\
  (n_L({\vec p}_3)+1)(n_L({\vec p}_4)+1).
\label{dG}
\ee
An event generator for this process has been constructed \cite{CP} where, in principle, one generates one event at a time continually updating the probability distribution in accordance
with the Bose enhancement factors.  Initially, all the occupation numbers
are zero but, after the first event, the next event is four times as likely
to be into the same phase space cell as into any other.  Because of the
huge number of available states, the second transition is still not
likely to be into the first phase space cell.  However, as soon as some
moderate fluctuation of selectrons has been produced in a single cell, 
the number in that state escalates rapidly producing a narrow jet of
selectrons.  As these decay down to the ground state via bremstrahlung,
a narrow jet of photons is created around the direction of the initial
selectron jet.  After a large number of events, a jet structure emerges as the distribution locks in on particular cells in the final state phase space.  To speed up the emergence of the jet structure, the occupation numbers are, in some studies, incremented by
two units at each event.  The distribution in momentum magnitude, polar angle cosine, and
azimuthal angle after $6 \cdot 10^{5}$ events is shown in table 1. The selectron
momentum will essentially be totally converted to gamma rays as the selectrons fall into the ground state although we have not as yet treated this radiative process in detail.

\vspace{0.2in}
\begin{table}[ht]
\begin{center}
\begin{tabular}{|rr|rr|rr|}\hline
 p (MeV)  &   N(p)   &   $\cos \theta$  & N($\cos \theta$)   &    $\phi$   &  N($\phi$) \\
\hline
 0.025 	  &    15    &	-0.90 	&   71969    & 	 0.314  &         43430   \\
 0.075 	  &   76     &  -0.70 	&  111973    &	 0.942  &	   70097  \\
 0.124 	 &    229    &	-0.50 	&  34399     &	 1.571  &    	  472570  \\
 0.174 	 &    553    &	-0.30 	&   19424    &   2.199  &	   34781  \\
 0.224 	 &   1000    &	-0.10 	&  632776    & 	 2.827  & 	  117413  \\
 0.274 	 &   1867    &	 0.10 	&  114240    &	 3.456  &  	  358925  \\
 0.324 	 &   2831    &	 0.30 	& 1272163    &	 4.084  &	   29715  \\
 0.373 	 &   9285    &	 0.50 	&  383288    &	 4.712  & 	 1813451  \\  
 0.423   &  107484   &	 0.70 	&   86555    &	 5.341  &	   26526  \\
 0.473 	 & 2886144   &	 0.90 	&  282697    &	 5.969  & 	   42576  \\
\hline
\end{tabular}
\caption{Selectron momentum and angular distributions showing the effect of boson enhancement}
\end{center}
\end{table}
\vspace{0.2in}

     Interesting features of the distribution are the jagged nature of the angular distribution and preference for momenta above the average of the Fermi sea.  This latter effect suggests that earlier (more probable) emissions will be of greater energy than later ones in agreement with indications from burst observations \cite{Pian}.

   There are obviously many details of the behavior of a dense
supersymmetric system remaining to be investigated but the SUSY
phase transition picture provides an alternative framework for thinking about gamma ray bursts that predicts some of the basic
characteristics and addresses some of the conceptual gaps of the
standard approaches. 
    
{\bf Acknowledgements}

    This work was supported in part by the US Department of Energy under grant DE-FG02-96ER-40967.  The work described here was done in collaboration with George Karatheodoris and Irina Perevalova.
In addition we gratefully acknowledge discussions with Doug Leonard, 
Phil Hardee, and Yongjoo Ou.

\end{document}